# Excitation and reception of pure shear horizontal waves by using face-shear $d_{24}$ mode piezoelectric wafers

Hongchen Miao[1,2], Qiang Huan[1], Faxin Li[1,2,a]

[1]LTCS and Department of Mechanics and Engineering Science, College of Engineering, Peking University, Beijing, 100871, China

[2]Center for Applied Physics and Technology, Peking University, Beijing, 100871, China

**Abstract**

The fundamental shear horizontal (SH0) wave in plate-like structures is of great importance in non-destructive testing (NDT) and structural health monitoring (SHM) as it is non-dispersive, while excitation and reception of SH0 waves using piezoelectrics is always a challenge. In this work, we firstly demonstrate via finite element simulations that face-shear piezoelectrics is superior to thickness-shear piezoelectrics in driving SH waves. Next, by using a newly defined face-shear $d_{24}$ PZT wafer as actuator and face-shear $d_{36}$ PMN-PT wafers as sensors, pure SH0 wave was successfully excited in an aluminum plate from 140 kHz to 190 kHz. Then, it was shown that the face-shear $d_{24}$ PZT wafer could receive the SH0 wave only and filter the Lamb waves over a wide frequency range (120 kHz to 230 kHz). The directionality of the excited SH0 wave was also investigated using face-shear $d_{24}$ PZT wafers as both actuators and sensors. Results show that pure SH0 wave can be excited symmetrically along two orthogonal directions (0° and 90°) and the amplitude of the excited SH0 wave can keep over 90% of the maximum amplitude when the deviate angle is within 30°. This work could greatly promote the applications of SH waves in NDT and SHM.

---

[a] Author to whom all correspondence should be addressed, Email: lifaxin@pku.edu.cn

**1. Introduction**

Ultrasonic guided wave inspection is a very efficient method in non-destructive testing (NDT) and structural health monitoring (SHM)[1]. The fundamental shear horizontal (SH0) waves in plate-like structures and torsional [T(0,1)] waves in pipes are especially useful for long-distance NDT due to their non-dispersive characteristics thus simplifying the interpretation of signals[2, 3]. Furthermore, as there is no out-of-plane particle displacement in SH0 wave and T(0,1) wave, they will be less affected by the presence of surrounding media. In addition, there is no mode conversion when these waves travel across defects or boundaries, e.g., SH0 wave may change to higher mode SH waves but will not convert to Lamb waves when encountering defects. Actually, the SH0 wave in plates is essentially equivalent to the T(0,1) wave in a thin pipe-like structure[4], so in most cases the conclusions obtained from one can be extended to the other straightforwardly.

Currently, although several methods have been proposed to generate SH0 or T(0,1) waves, it is still very difficult to obtain pure SH0 wave or T(0,1) wave mode efficiently. The well-known electromagnetic acoustic transducer (EMAT) can generate pure SH0 wave mode[5, 6], but EMATs can only work on conductive materials and the signal-to-noise ratio (SNR) is usually not high because of weak coupling with the structures due to its non-contact. Liu et al. employed large (12.5mm×8mm×4.5mm) thickness-shear ($d_{15}$) mode piezoelectric transducers to generate T(0,1) wave in pipes[7]. Recently, Kamal et al. used thickness-shear ($d_{15}$) mode piezoelectric wafers to generate SH0 wave in plates perpendicular to the poling direction, but strong lamb waves will be excited simultaneously along the poling direction[8]. Later, Boivin et al. showed that the amplitude of the co-excited Lamb waves can be decreased by optimizing the geometry of the $d_{15}$ mode piezoelectric wafer at a given frequency[9]. It should be noted that the energy conversion efficiency of thickness-shear ($d_{15}$) mode piezoelectric transducers is fairly low and the amplitude of the obtained SH0 wave is usually very small[8]. Zhou et al. used a face-shear $d_{36}$ mode PMN-PT crystal transducer to excite and receive SH0 waves in plates[10, 11]. Similarly, Lamb waves were also excited, since the extensional $d_{31}$ mode co-existed with the face-shear $d_{36}$ mode in PMN-PT crystals[12]. Moreover, the low Curie temperature and high cost of PMN-PT crystal make it not suitable for applications in NDT or SHM. Recently, we realized the face-shear $d_{36}$ mode in $PbZr_{1-x}Ti_xO_3$ (PZT) ceramics and successfully excited SH0 wave in an aluminum plate using $d_{36}$ mode PZT wafers[13-15]. Since the extensional $d_{31}$ mode also existed in the $d_{36}$ mode PZT ceramics, pure SH0 wave still cannot be obtained[15].

In this work, a newly defined face-shear $d_{24}$ PZT wafer (in-plane poled and electric field applied along another orthogonal in-plane direction) was proposed to excite and receive pure SH0 wave. Firstly, we demonstrate via finite element simulations that face-shear piezoelectric transducers are superior to thickness-shear ones in exciting SH waves in plates. Then we successfully excited pure SH0 waves and received the SH0 wave only in an aluminum plate using the face-shear $d_{24}$ mode PZT wafers as actuators or sensors. Finally, the directionality of the excited SH0 wave was investigated using the $d_{24}$ mode PZT wafers as both actuators and sensors.

## 2. Methodology

The first step in developing a SH wave transducer is to understand the excitation mechanism. It is well known that there exist two different families of guided waves in plates, i.e., Lamb waves and shear horizontal (SH) waves. The particle vibration caused by all the SH modes are in a plane parallel to the plate surface, and SH waves propagate perpendicular to the direction of particle's vibration[1]. Therefore, the thickness-shear $d_{15}$ mode transducer is a straightforward solution, since it may cause particle vibration in the x direction and SH waves can be excited in the z direction, as shown in Fig. 1(a). However, the energy conversion efficiency of the thickness-shear $d_{15}$ transducer is very low. Fig. 1(b) shows the finite element (FEM) simulated deformation of a thickness-shear $d_{15}$ PZT-5H wafer with dimensions of $6\,\text{mm} \times 6\,\text{mm} \times 1\,\text{mm}$ bonded on a 1 mm-thick aluminum plate through a thin bond layer. The thickness of the bond layer is set as 50 μm and its elastic modulus is 1 GPa. It can be seen that when bonded on an aluminum plate, the deformation of the thickness-shear $d_{15}$ wafer under a dc voltage (20 V) trends to be the simple shear deformation. i.e., the displacement of the bottom surface of the transducer is nearly zero and almost no deformation was transformed to the aluminum plate. Therefore, the SH waves in the plate can only be caused by the inertia force and the non-linear deformation of the thickness-shear $d_{15}$ piezoelectric wafer under AC voltages, resulting in a relative low efficiency of energy transfer.

Alternatively, inducing face-shear deformation in plates is another solution to generate SH waves,

which is the excitation mechanism of the $d_{36}$ mode piezoelectric transducers (crystals or ceramics). As mentioned above, the field induced deformation of the $d_{36}$ mode piezoelectric transducers is always the superposition of an extensional deformation and a pure face-shear deformation, since the extensional $d_{31}$ mode co-existed with the $d_{36}$ mode in $d_{36}$ type piezoelectric crystals and ceramics[12, 13]. Therefore, it is very difficult to excite pure SH wave by using the $d_{36}$ type piezoelectric transducers[11, 15]. Here a newly defined face-shear $d_{24}$ mode piezoelectric transducer was proposed to induce pure face-shear deformation in plates. As shown in Fig. 1(c), the PZT based face-shear $d_{24}$ wafer was in-plane poled along the "3" direction. The driving field is applied along another in-plane direction, i.e., the "2" direction and the thickness is along the "1" direction. Pure face-shear deformation is expected to be obtained in this type transducer under applied voltage, since $d_{24}$ mode is the unique deformation mode when the field is applied along the "2" direction. Compared with the conventional thickness-shear $d_{15}$ PZT transducer, the proposed face-shear $d_{24}$ PZT transducer is more efficient in driving SH waves in plates. Fig. 1(d) shows the FEM simulated deformation of a face-shear $d_{24}$ wafer ($6 \text{ mm} \times 6 \text{ mm} \times 1 \text{ mm}$) bonded on a 1 mm-thick aluminum plate under a dc voltage of 20 V. It can be seen that the face-shear deformation of the transducer can be effectively transferred to the plate through the bonding layer (the in-plane displacement transfer ratio is over 60%). The face-shear deformation will then cause the traction force $\tau_{xz}$ in the plate and the SH waves will be launched in both z direction and x direction, as indicated in Fig. 1(c).

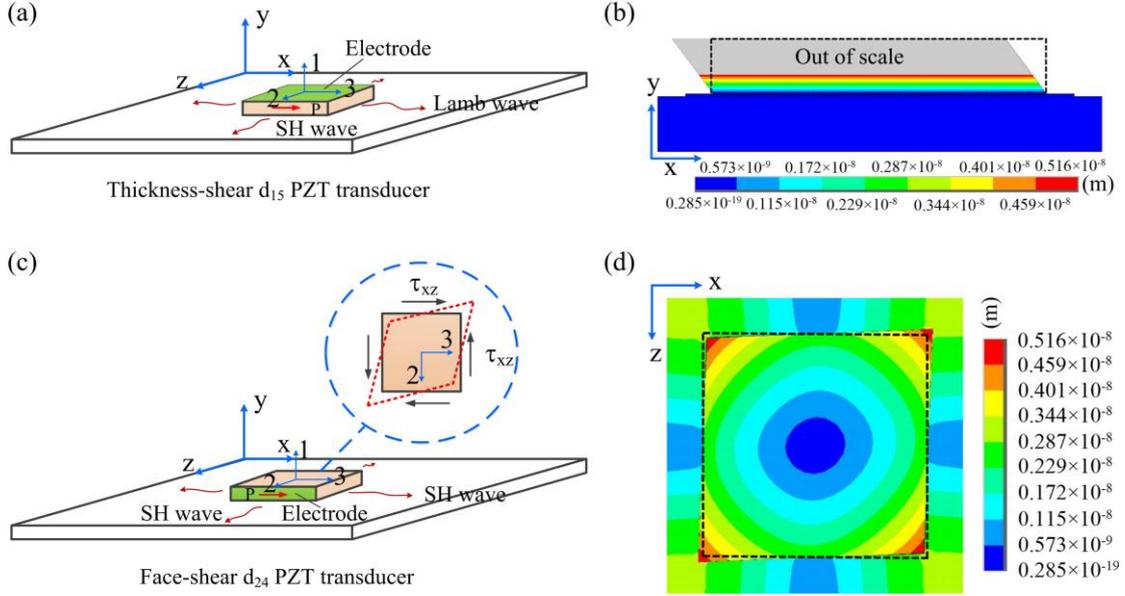

Fig. 1 Schematics of (a) the conventional thickness-shear $d_{15}$ PZT transducer and (c) the face-shear $d_{24}$ PZT transducer boned on a plate. FEM simulated in-plane deformations of (a) and (c) are shown in (b) and (d), respectively.

Experiments were then conducted to explore the face-shear $d_{24}$ PZT transducer's performance on excitation and reception of SH0 wave. An aluminum plate with the dimensions of $1000 \text{ mm} \times 1000 \text{ mm} \times 1 \text{ mm}$ was used in this study. The commercially available PZT-5H piezoelectric ceramics were used to fabricate the face-shear $d_{24}$ PZT wafer with dimensions of $6 \text{ mm} \times 6 \text{ mm} \times 1.5 \text{ mm}$. The $d_{24}$ wafer exhibits good face-shear performance from 130 kHz to 230 kHz, which was confirmed by its impedance spectrum as shown in Fig. 2(a). In the wave excitation/reception testing, firstly the face-shear $d_{24}$ PZT wafer served as the actuator to excite SH0 wave. A $d_{36}$ type PMN-PT crystal patch($d_{36}$=1600 pC/N and $d_{31}$=-360 pC/N) with dimensions of $5 \text{ mm} \times 5 \text{ mm} \times 1 \text{ mm}$ was used as the sensor to check the purity of the excited SH wave, since it can receive both SH wave and Lamb waves[11]. Meanwhile, a $d_{31}$ type PZT-5H sensor ($6 \text{ mm} \times 6 \text{ mm} \times 1 \text{ mm}$), which can only receive Lamb waves, was also used to detect the possible A0 and S0 waves. Then, to check whether the $d_{24}$ PZT transducer could only receive SH wave, a $d_{36}$ type PZT-4 wafer ($6.3 \text{ mm} \times 6.3 \text{ mm} \times 1 \text{ mm}$) which can generate both SH waves and Lamb waves, was used as the actuator. Detailed properties of the $d_{36}$ type PZT-4 patch can be found in our recent work [15]. Besides the $d_{24}$ PZT wafer, a $d_{36}$ type PMN-PT crystal wafer ($5 \text{ mm} \times 5 \text{ mm} \times 1 \text{ mm}$) was also used as sensor for comparison. Finally, the face-shear $d_{24}$ PZT wafers were used as both actuator and sensors to check the directionality of the excited SH0 wave. The layout and location of the actuator and sensors were shown in Fig. 2(b). All the actuators were driven by using a function generator (33220A, Agilent, USA), into which a five-cycle Hanning window-modulated sinusoid toneburst was preprogrammed. A power amplifier (KH7602M) was used to amplify the drive signal and an Agilent DSO-X 3024A oscilloscope was used to collect the wave signals received by sensors.

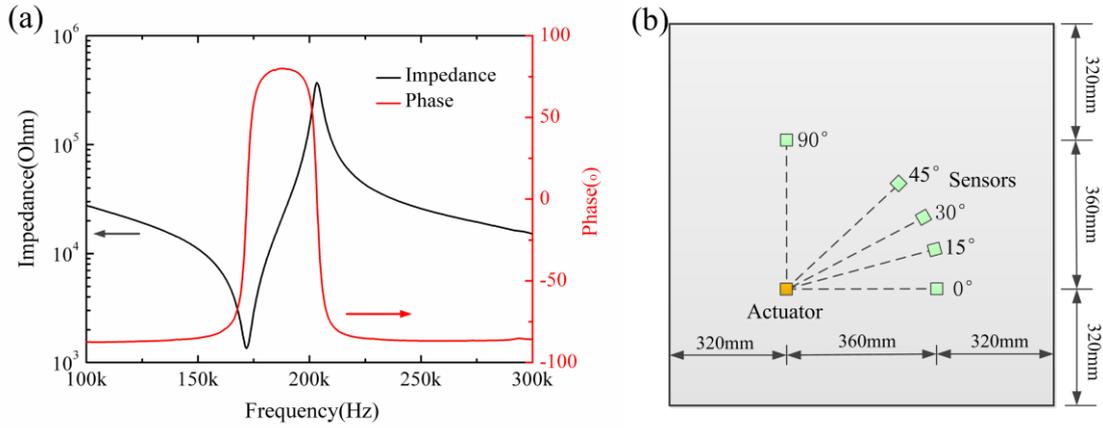

Fig. 2 (a) The impedance spectrum of the free face-shear $d_{24}$ PZT wafer with dimensions of $6~\text{mm}\times6~\text{mm}\times1.5~\text{mm}$. (b) Schematic of the layout of the piezoelectric actuator and sensors for wave excitation and reception in an aluminum plate.

## 3. Results and Discussions

Fig. 3(a) shows the wave signals excited by the face-shear $d_{24}$ PZT actuator and received by a $d_{36}$ type PMN-PT sensor placed on $0\degree$ direction. The central frequency of the drive signal was first fixed at 150 kHz and its amplitude was set to be 20 V. It can be seen from Fig. 3(a) that only SH0 wave with high signal-to-noise ratio (SNR) was detected and no other unexpected wave modes (such as S0 or A0 wave mode) were observed. As expected, the waveform of the received SH0 wave is almost the same as that of the incident pulse signal, i.e., no dispersion was observed. By using continuous wavelet transform (CWT) to analyze these signals, we can extract the time interval of 120.56 μs between the actuator and the sensor with the distance of 360 mm, as shown in Fig. 3(b). The calculated group velocity of 2986 m/s is in good agreement with the theoretical group velocity (3099 m/s) of SH0 wave in the aluminum plate. Moreover, the vanishing signals received by the $d_{31}$ type PZT sensor indicates that no S0 wave or A0 wave were generated, as shown in Fig. 3(c). Therefore, the testing results in Fig. 3 had verified that the proposed face-shear $d_{24}$ PZT wafer can excite pure SH0 wave at 150 kHz.

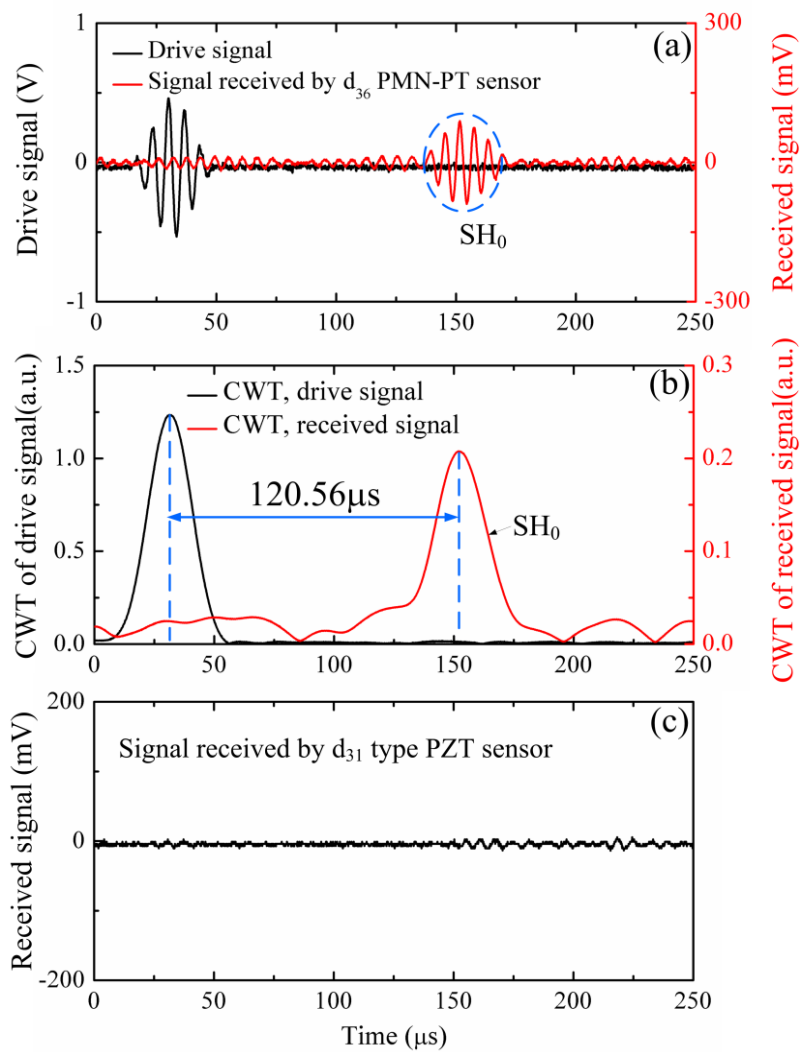

Fig. 3 Driving wave signals applied to the face-shear $d_{24}$ PZT actuator at 150kHz and received signals by (a) a $d_{36}$ type PMN-PT crystal sensor and (c) a $d_{31}$ type PZT sensor placed at 0° direction. (b) Continuous wavelet transform (CWT) of the driving and received wave signals shown in Fig. 3(a).

Fig. 4 shows the wave signals excited by the face-shear $d_{24}$ PZT actuator at different central frequencies and received by a $d_{36}$ type PMN-PT sensor placed on 0° direction. It can be seen that pure SH0 wave with high SNR can be excited by the face-shear $d_{24}$ PZT wafer over a wide frequency range (140 kHz to 190 kHz). Fig. 4(a) shows that pure SH0 wave can also be excited by the $d_{24}$ wafer at 130 kHz. However, the amplitude of the excited SH0 wave is relative small, since the exciting frequency is far away from the resonance frequency of the $d_{24}$ mode (192kHz for the bonded $d_{24}$ PZT wafer). When the exciting frequency was increased to 200 kHz, Fig. 4(g) shows that the S0 wave was also excited, which was further confirmed by wave signals detected by a $d_{31}$ PZT sensor, as shown in Fig. 4(h). Note that the frequency range from 140 kHz to 190 kHz is of practical significance, since frequencies in the range of 10kHz to 250 kHz are often used in engineering [16-18]. Also as expected, the waveform and group velocity of the excited SH0 wave are frequency-independent. On the other hand, it can be seen from Fig. 4 that the amplitude of the excited SH0 wave increases steadily with the increasing drive frequency, which may attribute to the fact that the face-shear deformation of the $d_{24}$ PZT wafer was amplified near the resonance frequency and the size of the $d_{24}$ PZT wafer became more and more close to the half of wavelength with the increasing drive frequency.

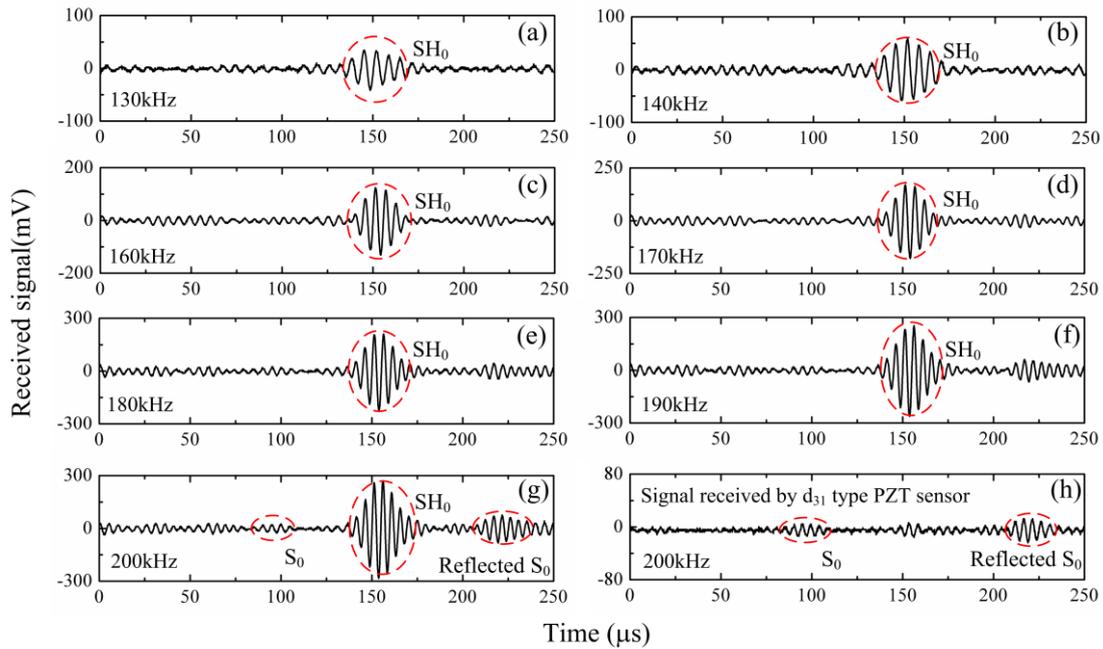

Fig. 4 Wave signals excited by the face-shear $d_{24}$ PZT actuator at different frequencies and received by a $d_{36}$ type PMN-PT crystal sensor ((a)-(g)) placed at 0° direction. (h) Wave signals excited by the $d_{24}$ PZT actuator at 200 kHz and received by a $d_{31}$ PZT sensor placed at 0° direction.

Then the face-shear $d_{24}$ PZT wafer served as a sensor to check its performance on receiving the SH0 wave. Here a $d_{36}$ type PZT-4 wafer was used as the actuator to excite both SH and Lamb waves. For comparison, a $d_{36}$ type PMN-PT crystal was also used as the sensor to receive the excited waves. It can be seen from Fig. 5(a) that both SH0 wave and A0 wave were successfully excited by using the $d_{36}$ type PZT-4 wafer under a voltage of 40 V at 120 kHz. However, when the excited waves were received by the $d_{24}$ PZT sensor, only SH0 wave was detected, as shown in Fig. 5(b). When the driving frequency varies from 120 kHz to 170 kHz, the same phenomenon was observed and here only the results at 170 kHz were given, as shown in Fig. 5(c) and (d). In addition, when the central frequency of the excitation signal increased from 170 kHz to 230 kHz, both lamb waves (A0 and/or S0 wave) and SH0 wave were excited simultaneously by the $d_{36}$ type PZT-4 actuator, and all these wave modes can be detected by the $d_{36}$ PMN-PT sensor. Again, when these waves were detected by the $d_{24}$ PZT sensor, only the SH0 wave was received and the lamb waves (A0 or S0 wave) were excluded. For simplicity, only the results at 220 kHz were given, as shown in Fig. 5(e) and (f). However, when the driving frequency increased to 240 kHz, Fig. (h) shows that the $d_{24}$ PZT sensor cannot filter the S0 wave reflected from the plate boundary. All these experiments indicate that the face-shear $d_{24}$ PZT wafer can also be used as a selective sensor, which can only receive SH wave and exclude the Lamb waves over a wide frequency range (120 kHz to 230 kHz in this work). Moreover, the wave form of the SH0 wave received by the $d_{24}$ PZT sensor is more perfect than that received by the $d_{36}$ type PMN-PT sensor, as shown in Fig. 5. To the author's knowledge, so far no sensor has been reported that can only receive the SH wave and exclude the lamb waves in a wide frequency range.

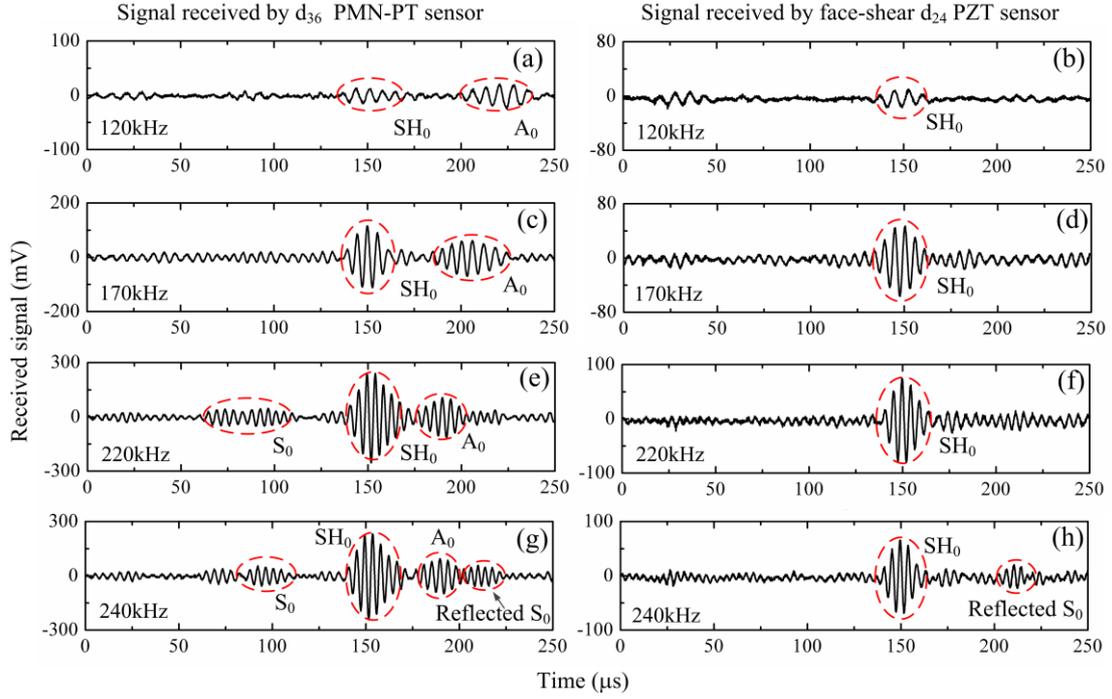

Fig. 5 Wave signals excited by a $d_{36}$ type PZT-4 actuator at different frequencies and received by a $d_{36}$ type PMN-PT sensor (left column) and by the face-shear $d_{24}$ PZT sensor (right column) placed at 0° direction.

The directionality of the SH0 wave excited by the $d_{24}$ PZT transducer was also investigated. Fig. 6(a) and (b) show the FEM simulated displacement wavefields excited by the $d_{24}$ PZT transducer bonded on a 1 mm-thick aluminum plate under a voltage of 20 V at 150 kHz. As we know, the SH waves propagate perpendicular to the direction of particle vibration, so the tangential displacement component in the cylindrical coordinate represents the SH0 wave, as shown in Fig. 6(a). It can be seen that the SH0 wave was excited along two orthogonal directions (0° and 90°) which can be defined as the main directions. Moreover, Fig. 6(a) shows that the excited SH0 wave is symmetric along the 45° direction. As indicted in our recent work[15], the pure face-shear deformation ($S_6$) is equivalent to the superposition of an elongation (or contraction) deformation along the 45° direction ($S_{45°}$) and a contraction (elongation) deformation along the 135° direction ($S_{135°}$). Therefore, theoretically the Lamb waves could be excited along the 45° (or 135°) direction. The FEM simulated results show that the amplitude of Lamb waves along 45° (or 135°) direction is quite small, since the tangential displacement component occupies about 95% of the total displacement, as shown in Fig. 6(a) and (b).

Experiments were then conducted to examine the FEM simulated results. Fig. 6(c) shows the SH0 wave excited and received by the face-shear $d_{24}$ PZT wafer along different propagation directions. The central frequency of the drive signal was fixed at 150 kHz and its amplitude was set to be 20 V. As expected, pure SH0 wave were excited along two orthogonal main directions (0° and 90°), as shown in Fig. 6(c). Furthermore, when the deviate angle referring to the main direction (0° or 90°) is less than 30°, no pronounced amplitude change of the SH0 wave was observed. At 45° direction, almost no wave can be detected by the face-shear $d_{24}$ PZT sensor. A $d_{31}$ type PZT sensor was further placed at the 45° direction to detect the possible Lamb waves and the result was plotted in the last subfigure of Fig. 6(c). It can be seen that S0 and A0 wave modes were indeed detected along the 45° direction, but their amplitude is rather small and can be neglected.

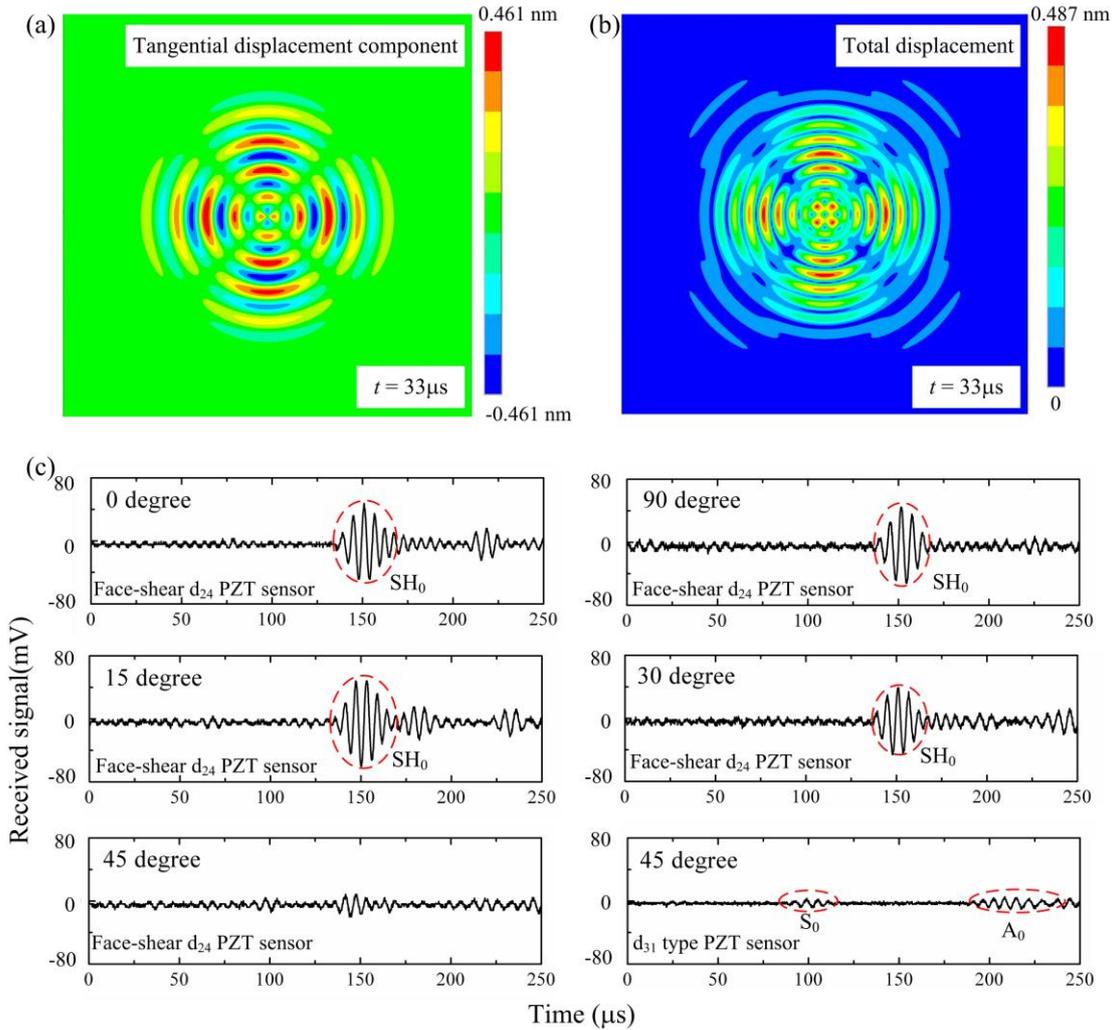

Fig. 6 FEM simulated displacement wavefields excited by the $d_{24}$ PZT transducer at 150 kHz: (a)

tangential displacement component (SH0 wave) in the cylindrical coordinate, (b) total displacement. (c) Experimental results: SH0 wave received by the face-shear $d_{24}$ PZT sensor or a $d_{31}$ PZT sensor at 150 kHz along different propagation directions.

## 4. Conclusions

In summary, we proposed a face-shear $d_{24}$ PZT wafer to excite and receive pure SH0 wave in plate structures. Finite element simulations indicate that the face-shear $d_{24}$ transducer is more effective than the thickness-shear $d_{15}$ transducer in driving SH waves. Experimental results show that pure SH0 wave with high signal-to-noise ratio (SNR) can be generated by the face-shear $d_{24}$ PZT wafer over a wide frequency range. Moreover, the $d_{24}$ PZT wafer can also serve as a selective SH wave sensor which can exclude lamb waves from 120 kHz to 230 kHz. The directionality of the excited SH0 wave was further investigated and it was found that pure SH0 wave can be excited along two orthogonal main directions (0° and 90°). Furthermore, the excited SH0 wave can keep its amplitude when the deviate angle referring to the main direction (0° or 90°) is less than 30°. The proposed face-shear $d_{24}$ PZT transducer could be of great importance in the fields of NDT and SHM, since it can excite/receive pure SH wave and be widely used as a cost-effective transducer.